\documentclass[%
 reprint,
 amsmath,amssymb,
 aps,
]{revtex4-2}

\DeclareUnicodeCharacter{2212}{-}
\usepackage{textgreek}
\usepackage{graphicx}
\usepackage{dcolumn}
\usepackage{bm}

\begin{document}

\preprint{APS/123-QED}

\title{Method for in-solution, high-throughput $T_1$ relaxometry using fluorescent nanodiamonds}
\author{Erin. S. Grant}
\affiliation{School of Physics, The University of Melbourne}
\author{Mina. Barzegar Amiri Olia}%
\affiliation{Bio21 Molecular Science and Biotechnology Institute}

\author{Ella. P. Walsh}
\affiliation{
 School of Chemistry, The University of Melbourne
}%

\author{Gawain McColl}
\affiliation{%
Florey Institute of Neuroscience and Mental Health
}%

\author{Liam T. Hall}
\affiliation{
 School of Chemistry, The University of Melbourne
}%

\author{David A. Simpson*}
\affiliation{School of Physics, The University of Melbourne}%
\email{simd@unimelb.edu.au}
\date{\today}

\begin{abstract}
Fluorescent nanodiamonds (FNDs) have been exploited as sensitive quantum probes for nanoscale chemical and biological sensing applications, with the majority of demonstrations to date relying on the detection of single FNDs. This places significant limits on the measurement time, throughput and statistical significance of a measured result as there is usually marked inhomogeneity within FND samples. Here we have developed a measurement platform that can report the $T_1$ spin relaxation time from a large ensemble of FNDs in solution.  We first describe a refined sensing protocol for this modality and then use it to identify the optimal FND size for the detection of paramagnetic targets. Our approach is simple to set up, robust and can be used for rapid material characterisation or a variety of \textit{in-situ} quantum sensing applications. 
\end{abstract}

\maketitle

\begin{figure*}
    \centering
    \includegraphics[width=\linewidth]{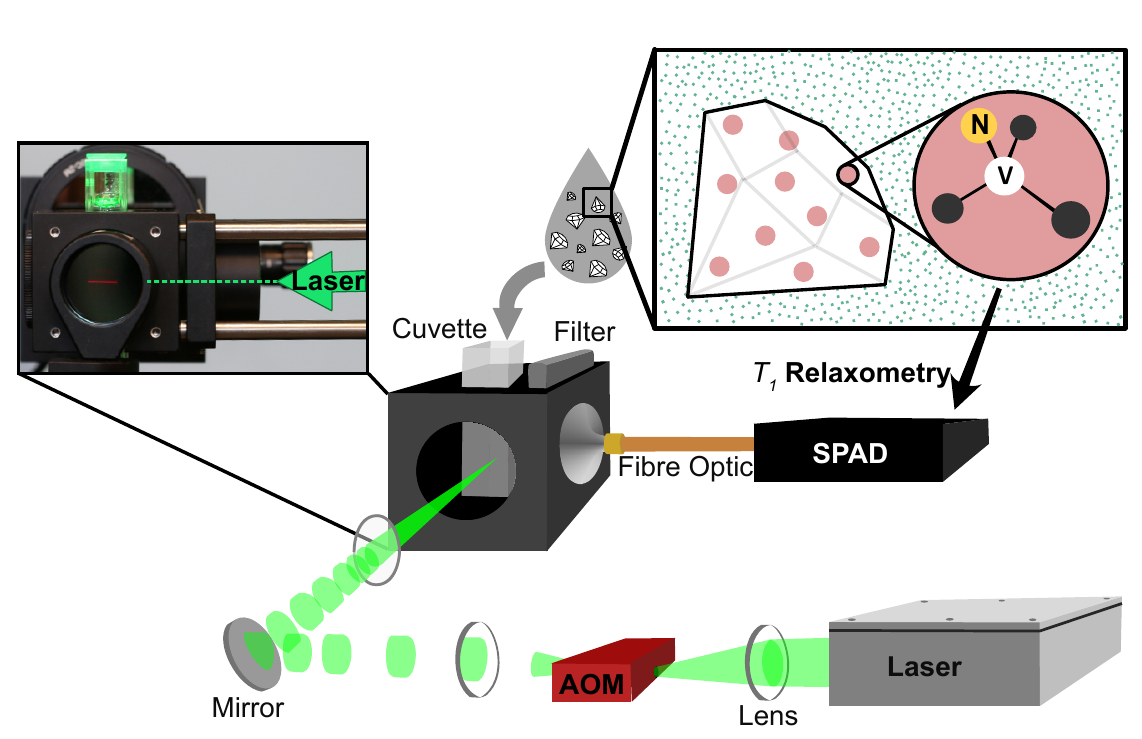}
    \caption{Schematic of the in-solution measurement apparatus. The set-up consists of a 532 nm laser which is modulated using an acousto-optical modulator. NV fluorescence is extracted through an optical fibre and measured on an single photon avalanche photodiode. Top left inset: A photo of the cuvette, NV fluorescence is visible as a thin red line through the optical filter at the front of the image. Top right inset: Quantum sensing is performed using dispersed FNDs containing the nitrogen-vacancy defect which consists of a substitutional nitrogen atom and an adjacent vacancy.}
    \label{fig:Apparatus}
\end{figure*}

Fluorescent nanodiamonds (FNDs) containing the nitrogen-vacancy (NV) defect have been developed over the last decade as a novel nanoscale sensor for a wide range of targets including pH \cite{Fujisaku2019, Fujisaku2019}, temperature \cite{Yukawa2020, Nishimura2021, Simpson2017, Tsai2017a}, free radicals \cite{Nie2021, Barton2020, Sharmin2021, Norouzi2022}, and paramagnetic targets such as manganese, gadolinium and iron \cite{Rendler2017a, McGuinness2013, Tetienne2013a, Ermakova2013, Akther2021}. The majority of characterisation and chemical sensing studies using FNDs address individual particles as they offer high sensitivity and spatial resolution  \cite{Reineck2019, Ermakova2013, Ryan2018, Nie2021, Padamati2022, Li2022}. However, these approaches are often limited in their statistical power due to the inhomogeneous nature of FND size, shape, brightness and coherence \cite{Reineck2019}. Therefore, alternate approaches which measure the average behaviour of an ensemble of FNDs offer robustness over individually sampling a small subset by reducing measurement bias. To date, chemical sensing using FND ensembles has been performed with particles deposited onto a 2D substrate. However, this approach can produce inhomogeneous coverage of the target across the FND sample, introducing large variabilities that reduce the quantitative utility of these measurements \cite{Gorrini2019a}. 

Here, we have established an all-optical, high-throughput protocol for reporting on the quantum properties of FNDs in-solution. This cuvette-based approach results in a more uniform interaction between FNDs and a target, facilitating \textit{in situ} measurement and characterisation using $T_1$ spin relaxometry. While this study focuses on the development of this novel modality for measuring the $T_1$ spin relaxation time of dispersed FNDs, this apparatus can also be modified to provide quantification of the average brightness and spectra of samples, as previously demonstrated \cite{Reineck2019a}.

This work contains two sections. First a protocol for $T_1$ spin relaxometry is established, which can account for fluctuations in the fluorescence intensity of the NV centres that are involved in a given measurement. We then apply this protocol to rapidly explore the best sized FNDs for paramagnetic sensing of chelated Gd$^{3+}$ ions in the form of gadobutrol - a chemical contrast agent used in magnetic resonance imaging (MRI). This proof of principle demonstration shows the simplicity and high-throughput of the technique as well as highlighting the ease of performing a robust control $T_1$ calibration measurement.

\section{\label{sec:level1} A Protocol for In-Solution $T_1$ Spin Relaxometry}

Measurements of dispersed FNDs were performed on a custom built set-up shown in Figure \ref{fig:Apparatus}. The apparatus consisted of a cuvette holder (Thorlabs), 532~nm (GEM, Laser Quantum) excitation laser,  acousto-optical modulator (AOM) (AA Electronics), and a single photon avalanche photodiode (SPAD) (Excelitas Technologies). Laser light of 160~mW was focused to the centre of the cuvette using a 150~mm plano convex lens, forming a focused beam with a waist of $\sim$100~\textmu m in diameter. NV fluorescence was extracted perpendicularly to the excitation beam through a  731$\pm$137~nm band pass filter. The NV fluorescence was coupled into an optical fibre (400~\textmu m core diameter) using an achromatic collimator with an NA of 0.54 (F950FC-A, Thorlabs) followed by detection with the SPAD. Micro cuvettes (BRAND 759200, Sigma Aldrich) were used to hold a 170~\textmu L suspension of FNDs. The left inset of Figure \ref{fig:Apparatus} shows a photo of the cuvette and holder. NV fluorescence is visible through the filter at the front of the holder as a red streak.
\begin{figure*}
    \centering
    \includegraphics[width=\linewidth]{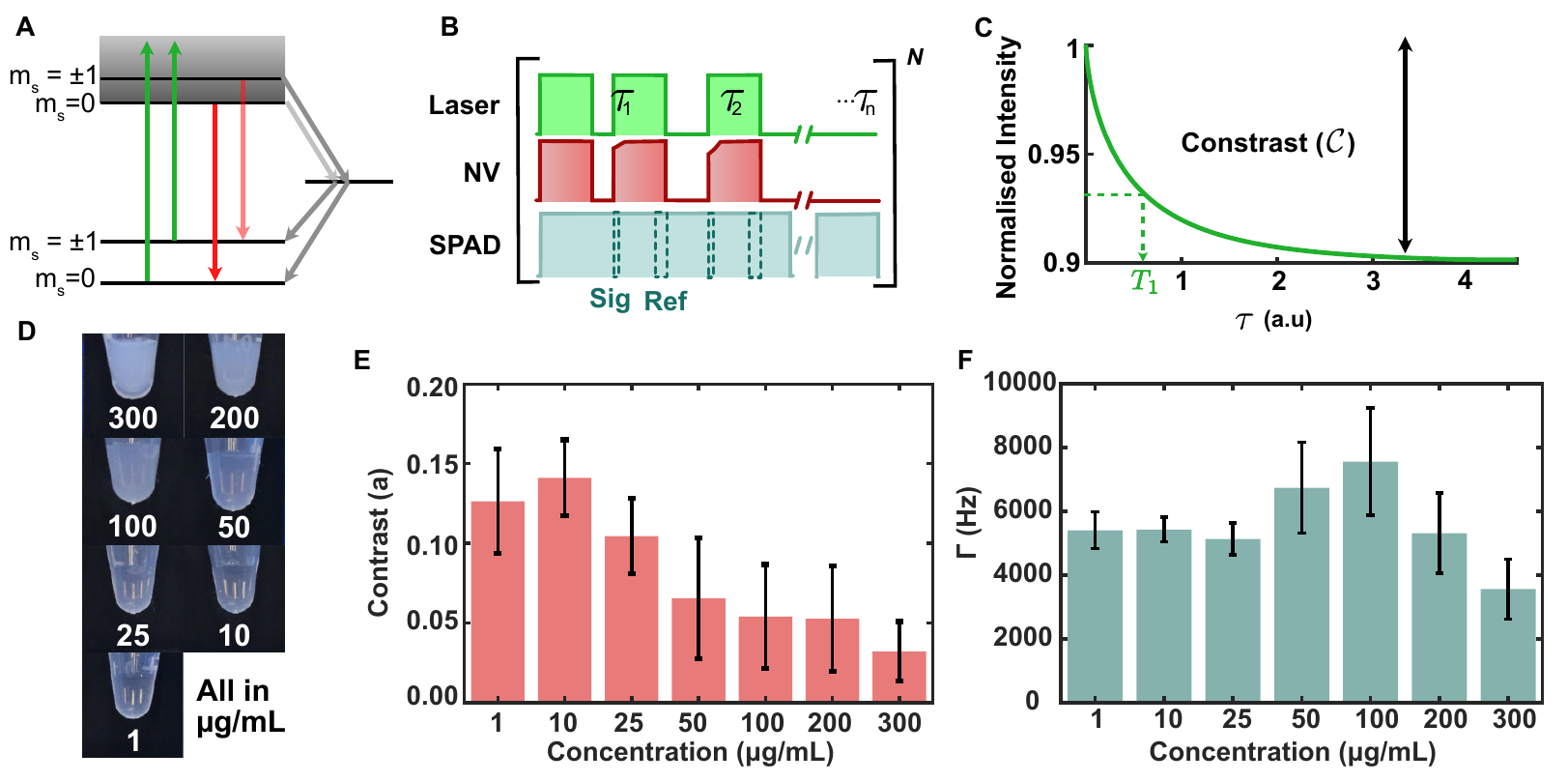}
    \caption{In-solution $T_1$ relaxometry with FNDs. A) Schematic of the energy levels of the NV defect in diamond. These consist of triplet ground and excited states and two intermediate singlet states. B) The NV spin population is determined by taking the ratio of the integrated intensity in the first 2 us of the excitation pulse by the integrated intensity in the final 10 us of the pulse. The signals are normalised to account for the differences in the integration window. C) A typical T1 relaxation curve measured using the pulse sequence in B). D) Images of 140 nm FND suspensions in ultrapure water at concentrations ranging from 1 to 300~\textmu g/ml. E) and F) The effect of the concentration of the FNDs on the measured $T_1$ spin contrast and $T_1$ relaxation time. The FND concentration has a negligible effect for concentrations below 50~\textmu g/mL.}
    \label{fig:NV}
\end{figure*}

$T_1$ spin relaxation measurements are facilitated by the intrinsic NV defects in the FNDs. The NV defect consists of a substitutional nitrogen atom and an adjacent vacancy, see the right inset of Figure \ref{fig:Apparatus}. It is a spin-1 system, with an electronic structure shown in Figure \ref{fig:NV}A. The $m_s=0$ and $m_s=\pm 1$ ground spin state levels are separated by the zero-field splitting, $D=2.87$~GHz. When excited from the $m_s=0$ ground state, the NV defect emits 30\% more fluorescence than the corresponding transitions from the $m_s=\pm1$ states which provides a convenient method to optically report the ground spin state populations \cite{Hopper2018, Jayakumar2018}. This can be used to perform $T_1$ spin relaxometry which measures the degree of longitudinal relaxation of the NV spin polarisation, where $T_1$ is defined as the time at which the emission intensity reaches $1/e$ of the maximum value. The characteristic time may also be expressed as a rate, $\Gamma = 1/T_1$, which will be used here. Fluctuating magnetic field sources act to increase the $T_1$ spin relaxation rate. These magnetic field sources can come from electronic spins on or within the FND \cite{Ryan2018, Laraoui2012}, or from external targets such as paramagnetic molecules \cite{Tetienne2013} or magnetic materials \cite{McCoey2020}. 

To measure the $T_1$ spin relaxation process, a simple protocol was employed, as shown in Figure \ref{fig:NV}B. First, the polarising laser was used to initialise the ensemble of NV defects within the beam path into the bright, $m_s=0$, ground state. Then, after an evolution time, $\tau$, a second laser pulse was used to read out the average spin-state of the NV ensemble, as well as repolarising for the next measurement. Each laser pulse in the protocol was applied for the same length of time, but by varying $\tau$, a $T_1$ spin relaxation curve of the form shown in Figure \ref{fig:NV}C can be generated. 

For an individual NV defect, the decay profile of the measured emission intensity will be determined by its intrinsic relaxation rate, $\Gamma_{\rm intrinsic}$, as well as that produced by the presence of $N_T$ target spins, $\Gamma_{\rm target}$. Giving emission intensity:
\begin{equation}
    I(\tau) = \mathcal{C}\exp[-\Gamma_{\rm intrinsic}\tau -\sum_j^{N_T}\Gamma^j_{\rm target}\tau]+c\\
\end{equation}

This is complicated further when considering an ensemble of FNDs, containing $N_{NV}$ NV defects in total: 
\begin{equation}
    I(\tau) = \sum_i^{N_{NV}}\mathcal{C}^i\exp[-\Gamma^i_{\rm intrinsic}\tau -(\sum_j^{N_T}\Gamma^j_{\rm target}\tau)^i]+c\\
\end{equation}
In this situation the emission intensity, $I(\tau)$, will deviate from a single exponential lineshape and the short time behaviour can be well approximated by fitting with a stretched exponential:
\begin{equation}
    \approx \mathcal{C}\exp[-(\Gamma_{\rm intrinsic}\tau)^p-(\Gamma_{\rm target}\tau)^q]+c
\end{equation}
where $\mathcal{C}$ is the spin contrast, $p$ and $q$ are stretch factors that are influenced by the variance in $\Gamma_{\rm intrinsic}$ and $\Gamma_{\rm target}$ across the FND ensemble, and $c$ is a constant offset. The distributions of both $\Gamma_{\rm intrinsic}$ and $\Gamma_{\rm target}$ depend on the geometric arrangement of the noise sources which contribute to each, i.e target spins in the case of $\Gamma_{\rm target}$ or intrinsic noise sources for $\Gamma_{\rm intrinsic}$. It is reasonable to assume that $p\approx q$ because $\Gamma_{\rm intrinsic}$ is dominated by  surface spin noise in FNDs \cite{Tetienne2013a, Casabianca2011, Laraoui2012}, while the influence of the target material will be skewed towards spins in close proximity to the surface due to the $r^{-6}$ dependence \cite{Tetienne2013a}. The fitted decay curve then becomes:
\begin{equation}
 I(\tau)=\mathcal{C}\exp[-(\Gamma_{\rm measured}\tau)^p]+c,
\end{equation}
where $\Gamma_{\rm measured} = \Gamma_{\text{intrinsic}} + \Gamma_{\text{target}}$, allowing the influence of the target to be isolated via:

\begin{equation}
\Gamma_{\text{target}} = \Gamma_{\text{measured}}-\Gamma_{\text{intrinsic}}.
\end{equation} 

To control for laser fluctuations, other environmental noise, and potential NV charge state changes across the measurement time, each data point in the $T_1$ curve is found by taking the ratio of the integrated emission intensity in a 2~\textmu s window at the beginning of each laser pulse (which informs on the average spin state population of the ensemble) with the integrated intensity in a 10~\textmu s window at the end of the pulse, which represents the steady state population of the repolarised state. The respective pulses are shown by the dotted boxes in the SPAD trace of Figure \ref{fig:NV}B. Each integrated intensity is normalised to account for the difference in width. 
\begin{figure}
    \centering
    \includegraphics[width=\linewidth]{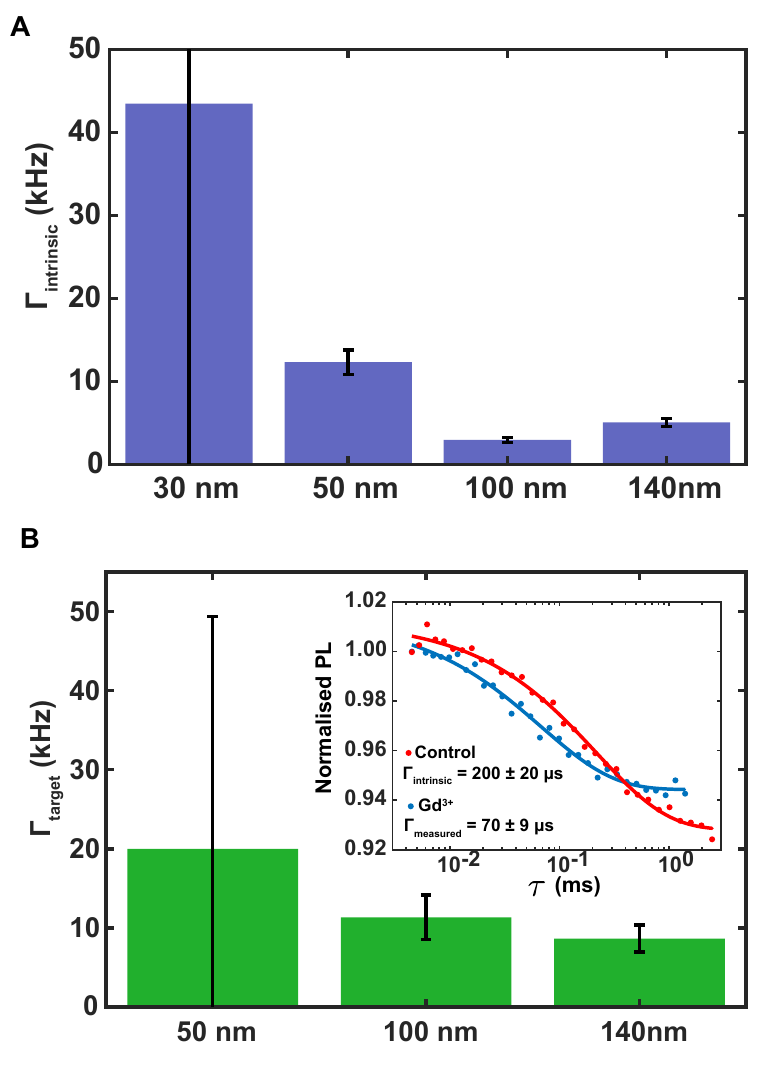}
    \caption{A) The intrinsic $T_1$ relaxation rate as a function of FND size. The error associated with the 30~nm sample has been truncated for clarity, it had $\Gamma_{\rm intrinsic}=40,000\pm70,000$. B) The target $T_1$ relaxation rate, $\Gamma_{\rm target}$, due to Gd$^{3+}$ for the 50, 100 and 140~nm samples. Inset: example curves for the control (red) and target (blue) measurements, using the 140~nm FNDs.}
    \label{fig:Gd}
\end{figure}

\section{ Proof of Principle Measurements}

To establish the feasibility of this new modality, initial $T_1$ relaxation measurements were performed on 140 nm FNDs purchased from Adamas Nanotechnologies and dispersed in ultrapure 18.2 M\textOmega resistant H$_2$O (Milli-Q, Millipore). The first consideration for this protocol was whether the low optical power density,  approximately three orders of magnitude less than traditional confocal microscopy, is sufficient for polarising the NV centres in the beam path. Secondly, it was important to understand whether charge state changes, which have been shown to cause artefacts to $T_1$ spin relaxation measurements \cite{Bluvstein2019}, are still important in this low power regime. To test these considerations we first measured the emission intensity over 100~\textmu s of continuous laser illumination to determine the polarisation timescale at this power. As shown in SI1, a steady-state is achieved after approximately 70-80~\textmu s of constant illumination. The optimal read out pulse length, $t_{RO}$, was then determined by measuring the steady-state intensity of each data point during a $T_1$ spin relaxation curve with different values of $t_{\rm RO}$. For the optimal readout pulse length of 80~\textmu s, the steady state population from each measurement point was within 2-3\% over the measurement period. To understand whether this small variation in steady-state emission intensity was caused by spin-independent processes such as NV charge state changes, we conducted a separate $T_1$ spin relaxation measurement under the influence of a strong external magnetic field (on the order of 0.1 T) \cite{Giri2019}. The introduction of a DC magnetic field induces spin-mixing, removing the ability to spin polarise the NV ensemble. Therefore, any observed variation in emission intensity, both in the reference window, as well as the signal window, should be spin-independent. As shown in SI2 the 2-3\% changes to the steady-state intensity are accounted for by spin-independent processes that appear to be occurring during the dark evolution times of the measurement, rather than within a read out pulse. Therefore, dividing the normalised intensity in the signal window by that of the reference window adequately functions as a common mode rejection method to account for these small charge state population changes \cite{Broadway2016}. 

After establishing a $T_1$ measurement protocol, we studied the effect of the FND concentration on the measured $T_1$ relaxation curve. Our findings show that as the concentration of the FND solution is increased, the NV spin contrast and signal to noise (SNR) of the measurement decreases. This is attributed to photon scattering at concentrations greater than 25~\textmu g/ml as illustrated in Figure \ref{fig:NV}E. The scattering effects also mask the $T_1$ relaxation effects, as seen in Figure \ref{fig:NV}F, leading to significant measurement errors. Therefore, we work with FND concentrations less than 25~\textmu g/ml for the remainder of this work. We note this concentration limit is likely to be a lower bound as scattering increases with particle volume \cite{Austin2020}. It should also be noted that efficacy at low FND concentrations is important for chemical sensing applications where the concentration of target molecules is low. For example, an over abundance of FNDs for a given target concentration will reduce the number of NVs in the ensemble that experience an additional relaxation process, $\Gamma_{\rm target}$. Therfore the measured relaxation rate, $\Gamma_{\rm measured}$, will be skewed towards intrinsic values, limiting SNR. 

\section{Paramagnetic Sensing in Solution}

In-solution measurement of $T_1$ relaxatometry has the potential to offer high-throughput measurements of large numbers of FNDs, making it useful for both characterisation and quantum sensing studies. Choosing a particle for quantum sensing purposes is often about balancing dynamic range, sensitivity, and the limit of detection (LOD). Ideally, a sensor will respond consistently to a wide range of concentrations with a low LOD, but it must also be able to distinguish between two samples with very similar concentrations. In the context of $T_1$ spin relaxometry, optimal performance within this metric will be determined by the intrinsic $T_1$ spin relaxation rate of the FNDs, their optical noise and the depth distribution of the NV defects within each FND. If all other characteristics are equal, FNDs with a lower intrinsic $T_1$ spin relaxation rate, $\Gamma_{\rm intrinsic}$, offer a greater dynamic range, as the operating band width will span $\Gamma_{\text{max}}-\Gamma_{\text{intrinsic}}$, where $\Gamma_{\rm max}$ is the maximum possible measured relaxation rate, which is determined by the spin polarisation time (on the order of microseconds). Meanwhile, the measurement sensitivity depends on the NV spin contrast, photon count rate, and the total acquisition time of the measurement. For a given evolution time, $\tau$, assuming shot-noise limited data, the signal-to-noise ratio of a $T_1$ spin relaxation measurement is given by:

\begin{equation}
    \text{SNR}(\tau)\approx\sqrt{\frac{\mathcal{R}t_{\rm RO}T_{\rm tot}}{\tau}}\frac{3\mathcal{C}}{4}e^{-\Gamma_{\rm intrinsic}\tau}(1-e^{-\Gamma_{\rm measured}\tau}),
\end{equation}
where $\mathcal{R}$ is the photon count rate, $t_{\rm RO}$ is the laser read-out length, $T_{\rm tot}$ is the total acquisition time, and $\mathcal{C}$ is the $T_1$ spin contrast \cite{Wood2016}. The protocol presented here is approximately a factor of 2 away from the shot noise limit when comparing the measured photon statistics with the expected shot noise limited value (see SI3). The additional optical noise may arise from photon scattering and diffusion of the FNDs themselves as well as residual laser intensity fluctuations not accounted for by the common mode rejection method. Nevertheless, we are able to determine the $T_1$ spin relaxation rates with measurement errors less than 10\% for a measurement time of $\sim$50 minutes. Finally, FNDs with the shallowest depth distributions will offer the lowest LOD, as the magnetic interaction between a single NV defect and a target molecule scales as $1/r^6$.

To explore the interplay between FND size, LOD, sensitivity, and dynamic range, we measured the intrinsic $T_1$ spin relaxation rates of four different FND sizes as well as their sensitivity to Gd$^{3+}$ ions. FNDs with a nominal diameter of 30, 50, 100 and 140~nm were sourced from Adamas Nanotechnologies. Particles were suspended in ultrapure water and sonicated for 30 seconds to ensure optimal dispersion. Initially, the intrinsic $T_1$ relaxation rate of each FND sample was measured using a 170~\textmu L volume.  Following this initial measurement, the sensitivity was determined by adding a 30~\textmu L solution of 26.6~mM of gadobutrol monohydrate (Sigma Aldrich). The Gd$^{3+}$ ion bound to this ligand has an electron spin of $S=7/2$, providing a fluctuating magnetic signal that increases the $T_1$ spin relaxation rate of the NV defect as shown by the example $T_1$ relaxation curves in the inset of Figure \ref{fig:Gd}B. To ensure a similar level of photon shot noise between measurements, data was acquired until $\sim$1.2 million photons were collected in the reference window of the first data point. The brightness of each FND sample varied, with the 30 and 50~nm samples producing <1 million counts per second. The 100 and 140 nm samples however, produced an order of magnitude more NV fluorescence. Therefore, to ensure the SPAD detector operated within its linear range, a neutral density filter with an optical density of 1.6 was added to reduce the count rate to <2 million counts/sec.

\begin{table}
\caption{Properties of each FND size and resulting SNR. The contrast value is taken from the intrinsic relaxation curve.}
\label{Tab:Results}
\begin{ruledtabular}
\begin{tabular}{lccr}
Size  & Brightness   & Contrast  & SNR\\
(nm) &  (mil photons/sec)  &  (\%) & \\
\hline
30       &   0.7    & 2 $\pm$ 8  & N/A \\
50        & 0.9  &  4 $\pm$ 1 & 16 $\pm$ 9\\
100       &  11  & 9 $\pm$ 2 & 43 $\pm$ 9 \\
140       & 13  & 10 $\pm$ 2 & 37 $\pm$ 10 \\
\end{tabular}
\end{ruledtabular}
\end{table}

Figure \ref{fig:Gd}A shows the intrinsic $T_1$ spin relaxation rate, $\Gamma_{\rm intrinsic}$, of each FND sample in ultrapure water, while Figure \ref{fig:Gd}B illustrates the effective $T_1$ spin relaxation rate, $\Gamma_{\rm target}$, resulting from the addition of Gd$^{3+}$ ions to the solution. The error bars represent one standard deviation of the error on the $T_1$ fitted parameter.

There is a relatively strong negative correlation between $\Gamma_{\rm intrinsic}$ and the size of the particle; as the size of the particle increases the intrinsic $T_1$ spin relaxation rate decreases. This may be related to the higher surface-area-to-volume ratio of the smaller particles, which increases the deleterious effects of unwanted surface spin noise. The 30~nm sample was especially affected, with an intrinsic $T_1$ spin relaxation rate significantly higher than any of the other samples and a prohibitively large fit error ($\Gamma_{\rm intrinsic}=40,000\pm70,000$). For this reason gadobutrol was not added to the 30~nm FNDs. 

Overall, the 100~nm FNDs outperformed the larger 140~nm particles, suggesting that the 100 nm sample has the highest dynamic range and greatest sensitivity. This is confirmed by the values of $\Gamma_{\rm target}$ recorded for each FND size and the corresponding SNR (calculated at $\tau = T_1/2$ for each particle), shown in Table 1. In contrast, $\Gamma_{\rm target}$ is much larger for the 50~nm FNDs compared to the 100 or 140~nm samples, suggesting they can offer a lower LOD. However, these particles also have a much higher measurement error, likely because the addition of gadobutrol increased $\Gamma_{\rm measured}$ to a value close to $\Gamma_{\rm max}$. At this level of perturbation the system cannot be as effectively spin polarised and hence there is a reduction in spin contrast (see SI6) which reduces the SNR. Additionally, the intrinsic relaxation rate of these particles is high, which means they have a much smaller dynamic range compared to the 100 and 140~nm FNDs. Importantly however, if $\Gamma_{\rm intrinstic}$ can be reduced for the smaller particles, they have the potential to sense very low concentrations of target material with a wider dynamic range due to their more favourable NV depth distribution. Nevertheless, in their current form, we haave identified the 100~nm FNDs to be the most attractive particles for paramagnetic quantum sensing.

This simple in-solution set-up has demonstrated the feasibility of measuring the average $T_1$ spin relaxation rate of dispersed FNDs. We note there is room to optimise the current configuration further. For example, the use of a SPAD for detecting the NV fluorescence offers highly time-resolved data, which was useful for validating the $T_1$ spin relaxation protocol, however it puts constraints on the amount of fluorescence that can be detected. Consequently, the acquisition speed is currently limited to measurements on the order of half an hour to an hour. Throughput could therefore be improved by utilising an alternative detector, such as a photodiode. For this to be feasible the collection efficiency must be further improved. Pre-detector elements to achieve this include aspheric lenses \cite{Webb2020}, integrating spheres \cite{Porres2006}, or parabolic lenses, which may provide up to 65\% collection efficiency \cite{Wolf2015}, substantially increasing acquisition speed.

In this work we have demonstrated that $T_1$ spin relaxometry can be performed on dispersed FNDs in an all-optical cuvette-based system, reporting from millions of particles simultaneously. We have shown how this system can be used for rapid assessment of the suitability of current commercially available FND materials for paramagnetic quantum sensing applications. From our measurements 100 nm FNDs currently offer the largest dynamic range and highest sensitivity for chemical sensing, however smaller particles have the potential to reach lower limits of detection. This measurement platform will also be useful for understanding the effects of surface functionalisation on $T_1$ spin relaxation times \cite{Wolcott2014, Stacey2019} - providing a method for assessing the impacts of a particular termination with high statistical power and throughput, enabling more streamlined advancement to FND materials. Finally, the in-solution nature of the measurement allows for \textit{in-situ} chemical sensing which would take orders of magnitude longer if pursued using standard confocal or widefield microscopes. With straight forward improvements to the collection efficiency, or by implementing a single point $T_1$ protocol, this system can be improved further, with the potential to offer sub second acquisition speeds. This \textit{in situ} technique opens the door to new sensing paradigms for these nanoscale fluorescent quantum sensors providing a path to well controlled, time-resolved measurements of dynamic chemical processes. 

\section{Supplementary Material}

See Supplementary Material for further information on how the all-optical protocol was optimised.

\section{Acknowledgements}

The authors would like to acknowledge Dr P. Reineck for supplying the nanodiamond material used in this study. E.G acknowledges the Graeme Clark Institute for Biomedical Engineering at the University of Melbourne for funding support. 

\section*{Data Availability Statement}
The data that support the findings of this study are openly available in [repository name] at http://doi.org/[doi], reference number [reference number].

\bibliography{library}

\end{document}